\documentclass[]{spie}

\usepackage{amsmath,amsfonts,amssymb}
\usepackage{graphicx}
\usepackage{xcolor}
\usepackage[normalem]{ulem}
\usepackage{float}
\usepackage[colorlinks=true, allcolors=blue]{hyperref}

\title{Engineering a Phase-Noise-Based Quantum Random Number Generator for Real-Time Secure Applications: Design, Validation, and Scalability}

\author{Anurag K. S. V.}
\author{Shubham Chouhan}
\author{K. Srinivasan}
\author{G. Raghavan*}
\author{Kanaka Raju P.*}
\affil {School of Quantum Technology, Defence Institute of Advanced Technology (DIAT), Pune, MH-411025, India}

\authorinfo{Further author information: (Send correspondence to G.R.*, K.R.P.*)\\A.K.S.V.: E-mail: anurag.krovvidi@gmail.com\\S.C.: E-mail: sschouhan625@gmail.com\\K.S.: E-mail: srinik@diat.ac.in\\G.R.*: E-mail: graghavan@diat.ac.in\\K.R.P.*: E-mail: raju@diat.ac.in}

\begin{document} 
\maketitle

\begin{abstract}
    Random Number Generators (RNGs) are crucial for applications ranging from cryptography to simulations. Depending on the source of randomness, RNGs are classified into Pseudo-Random Number Generators (PRNGs), True Random Number Generators (TRNGs), and Quantum Random Number Generators (QRNGs). This work presents the end-to-end development of a high-speed, high-efficiency, phase-noise-based QRNG system that taps into the quantum phase noise of a single-frequency laser, with randomness originating from spontaneous emission. Using a self-heterodyne measurement with a semiconductor laser (linewidth $\approx$ 5.23 $GHz$) operated near threshold and a $\sim$48 $cm$ fiber delay line, a raw data generation rate of 2.0 $Gbps$ is achieved. To ensure uniform randomness in the QRNG output, robust extraction techniques developed in-house, such as the Toeplitz Strong Extractor (TSE), are used. Randomness validation using the NIST and Diehard test suites confirms that all statistical tests pass at standard confidence levels. The developed system achieves a post-processed generation rate of 1.0 $Gbps$ in operation and attains a Technology Readiness Level (TRL) of 7, approaching TRL 8, making it suitable for real-time secure applications such as cryptographic key generation and stochastic modeling.
\end{abstract}


\section{INTRODUCTION}
\label{sec:intro}

    RNGs are the primary source of randomness in many applications, including understanding weather patterns, predicting the stock market via Monte Carlo simulations, analyzing the random Brownian motion of molecules, generating keys, hash codes, and many more \cite{downey_2012_randomness, calude_2016_classical, kollmitzer_2020_quantum}. These RNGs can be classified based on their source of randomness into PRNGs, that rely on mathematical algorithms \cite{deng_2017_developments, matsumoto_1998_mersenne}; TRNGs, that draw from seemingly random physical processes \cite{barakat_2013_generalized, hamburg_2012_analysis}; and QRNGs, that exploit the inherent randomness associated with quantum mechanics \cite{kollmitzer_2020_quantum, prl_2024_investigatingdiqrng, cheng_2024_semideviceindependentqrng, lang_2024_onchipsourcediqrng, nie_2024_measurementdiqrng, tanizawa_2024_realtime, roman_2023_phasenoiseqrng}. 

    QRNGs, whose source of randomness stems from the inherent uncertainty associated with quantum processes, and hence can be deemed ontologically random\cite{kollmitzer_2020_quantum}. Rapid developments in the field of quantum technologies have accelerated the development of these devices. Especially, the discovery of Shor’s algorithm \cite{shor1994, shor_1997_algo} posed a threat to the current asymmetric key cryptography systems, which serve as a backbone to the current-age communication infrastructure over the internet, including but not limited to online banking, secure messaging, and high-security communication. To counter this disruptive technology, Quantum Key Distribution (QKD) systems \cite{xu_2020_qkd, mehic_2020_qkd, zhang_2024_cvqkd, kundu_2024_thzqkd, Zhuang_2025_QKD404, Zhang_2025_SPQKD, Yu_2025_timebinEntQKD, Li_2025_MicroSatelliteQKD, Tagliavacche_2025_EntQKD}, and Post-Quantum Cryptography (PQC) schemes \cite{bernstein_2017_pqc, kumar_2020_pqc, joseph_2022_pqc, dam_2023_pqc, ritik_2023_pqc, allgyer_2024_pqc, rawal_2024_pqc, Turnip_2025_PQC} are being developed, which are in principle 100\% resilient to any classical or quantum adversary’s passive attacks. A major subsystem of all QKD protocols is a QRNG, which serves as the quintessential source of randomness required for independent, unpredictable choices that need to be made as part of the communication protocol \cite{kumar_2021_stateoftheart, vasani_2024_qkdpqc}.

    \begin{figure} [ht]
       \begin{center}
            \begin{tabular}{c} 
                \includegraphics[height=5cm]{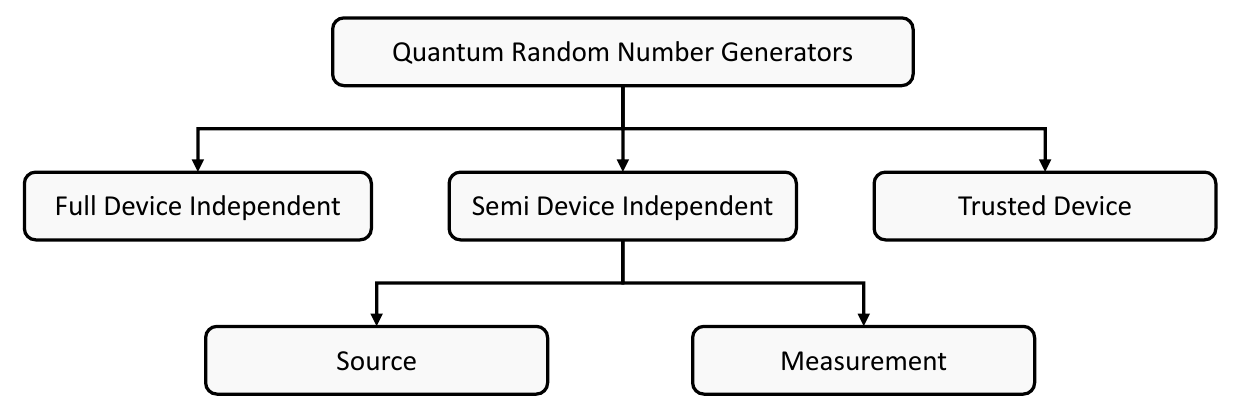}
            \end{tabular}
       \end{center}
       \caption[ParadigmsOfQRNG] 
       { \label{fig:ParadigmsOfQRNG} Various paradigms of QRNGs include fully device-independent QRNGs (FDI-QRNGs), semi-device-independent QRNGs (SDI-QRNGs), and trusted-device QRNGs (TD-QRNGs). SDI-QRNGs are further divided into source and measurement subgroups.}
       \end{figure} 

    These QRNGs can be further classified into Full Device Independent QRNGs (FDI-QRNGs), Semi-Device Independent QRNGs (SDI-QRNGs), and Trusted Device QRNGs (TD-QRNGs), as shown in Fig.~\ref{fig:ParadigmsOfQRNG}. FDI-QRNGs aim to generate random numbers whose unpredictability can be certified without trusting the devices used for generation and measurement \cite{liu_2018_diqrng, raghavan_2021_device}. The randomness is solely guaranteed by the violation of a Bell inequality \cite{dehlinger_2002_belltest, acn_2016_certified, yzha_2018_certquantrand}. Although FDI-QRNGs offer the highest level of security, they can be extremely challenging to implement experimentally \cite{liu_2018_diqrng, zhang_2020_diqr, raghavan_2021_device, prl_2024_investigatingdiqrng}, and their generation rates are quite low for any real-world application. Alternatives to FDI-QRNGs are SDI-QRNGs, which offer better efficiency as compared to FDI-QRNGs with reasonable device assumptions. These SDI-QRNGs come in two forms: (a) Source-SDI-QRNGs \cite{marangon_2017_sourcediqrng, avesani_2018_sourcediqrng, cheng_2022_mutuallysourcediqrng, Lin_2022_imperfqrng, lang_2024_onchipsourcediqrng, Du_2025_sourcesdiqrng} and (b) Measurement-SDI-QRNGs \cite{nie_2024_measurementdiqrng, cao_2015_measurementdiqrng, nie_2016_measurementdiqrng, song_2019_measurementdiqrng, wutao_2023_measurementdiqrng}. 

    To overcome the stringent limitations posed by FDI/SDI-QRNGs, we move towards exploring TD-QRNGs. These QRNGs are known for their high generation rates suitable for industry applications, primarily in various QKD protocols\cite{xu_2020_qkd, mehic_2020_qkd, zhang_2024_cvqkd, kundu_2024_thzqkd, Zhuang_2025_QKD404, Zhang_2025_SPQKD, Yu_2025_timebinEntQKD, Li_2025_MicroSatelliteQKD, Tagliavacche_2025_EntQKD}. TD-QRNGs assume well-characterized quantum entropy sources and measurement devices to produce randomness. Their theoretical simplicity and ease of implementation make them practical and industry-ready. Trust in the device’s proper functioning allows the use of suitable quantum protocols to generate random numbers\cite{ap_2023_rngreview}. In order to distill out the quantum randomness and filter any influence by an adversary, most of these devices undergo a post-processing process involving randomness extraction\cite{ma_2013_postprocessing, kollmitzer_2020_quantum, guo_2024_parallelppqrng, ksv_2024_tsecpugpu, aksv_2024_tsefpga, Foreman_2025_cryptomite}.

    Initial TD-QRNGs involved using the quantum mechanical description of a beam-splitter, and the use of single photon detectors equispaced on the reflecting path and the transmitting path post the beam splitter. Based on whether the photon gets reflected or transmitted, we constitute a click in the transmitted detector as binary zero and the reflected detector as binary one\cite{jennewein_2000_beamsplitterqrng}. While these systems are simple to construct, the quality of random number generation is limited by the balanced nature of the beam splitter and the efficiency of the single-photon detection systems. Owing to the engineering limitations of the beam-splitter-based systems, further developments were carried out towards using the time of arrival of photons, optical parametric oscillator-based TD-QRNGs\cite{sunada_2011_opoqrng, marandi_2012_opoqrng}, etc., while these systems showed improvements, the problem of high random number generation rates was still persistent.

    To address these challenges and make TD-QRNGs speed comparable for cryptographic applications and QKD systems, the community has turned towards using vacuum fluctuations as an entropy source. This is achieved by combining a quantum vacuum state (empty port) and a strong coherent beam into the two input ports of a balanced homodyne detector, thereby generating random numbers.\cite{huang_2019_homodyneqrng, gehring_2021_homodyneqrng, bai_2021_188, wang_2023_homodyneqrng, bru_2023_homodyneqrng, tanizawa_2024_realtime, qiao_2025_realtimeonchiphomodyneqrng}. Similarly, TD-QRNGs optimized for harvesting the laser’s phase noise have been realized\cite{qi_2010_highspeed, xu_2012_ultrafast, raffaelli_2018_phasenoiseqrng, lei_2020_phasenoiseqrng, roman_2023_phasenoiseqrng, Huang_2025_onchipphasenoiseqrng}. Both vacuum state and phase-noise-based TD-QRNGs require post-processing to generate information-theoretically provable random numbers\cite{ma_2013_postprocessing, gehring_2021_homodyneqrng, guo_2024_parallelppqrng}. The generation rates of the aforementioned TD-QRNGs are listed in Table~\ref{tab:tdqrng_rates}.

    \begin{table}[ht]
        \caption{Generation rates of various trusted-device quantum random number generators (TD-QRNGs).}
        \label{tab:tdqrng_rates}
        \begin{center}
            \begin{tabular}{|l|l|} 
                \hline
                \rule[-1ex]{0pt}{3.5ex}  Type & Rate (Order) \\
                \hline
                \rule[-1ex]{0pt}{3.5ex}  Optical Parametric Oscillators & $Kbps$ \\
                \hline
                \rule[-1ex]{0pt}{3.5ex}  Beam-Splitter & $Mbps$ \\
                \hline
                \rule[-1ex]{0pt}{3.5ex}  Vacuum Fluctuations & $Gbps$ \\
                \hline
                \rule[-1ex]{0pt}{3.5ex}  Phase Noise & $Gbps$ \\
                \hline
            \end{tabular}
        \end{center}
    \end{table}

\begin{figure} [ht]
       \begin{center}
            \begin{tabular}{c} 
                \includegraphics[height=4cm]{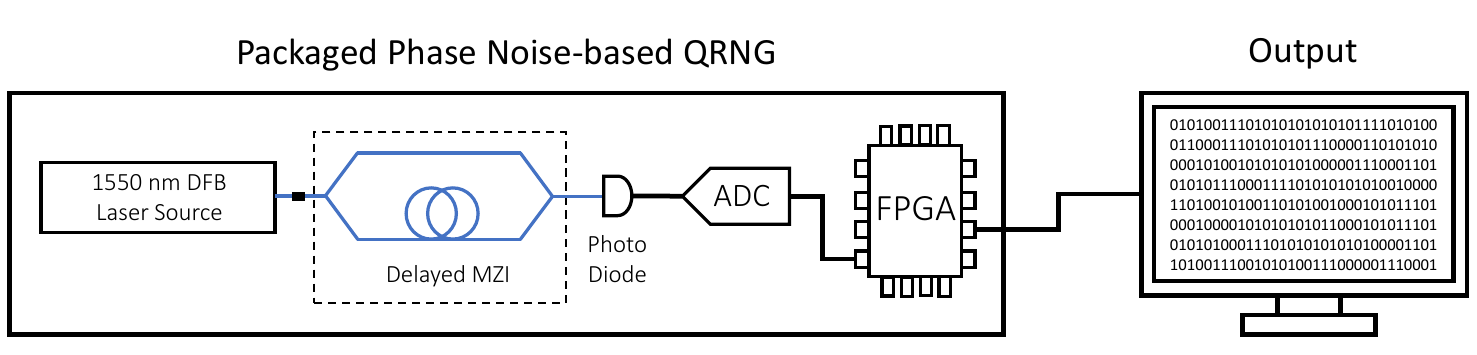}
            \end{tabular}
       \end{center}
       \caption[PackagedPhaseNoiseQRNG] 
       { \label{fig:PackagedPhaseNoiseQRNG} Packaged phase-noise-based QRNG schematic. Components include a distributed feedback (DFB) laser source, a delayed Mach-Zehnder interferometer, a photodiode, an analog-to-digital converter (ADC), and a field-programmable gate array (FPGA).}
       \end{figure}

    This work presents the development of one of the major realizations of a TD-QRNG based on harvesting the spontaneous emission of a laser source via delayed self-heterodyne phase-noise measurement. The schematic of the packaged form of the phase-noise-based QRNG is shown in Fig.~\ref{fig:PackagedPhaseNoiseQRNG}. The phase-noise-based QRNG offers hardware simplicity as compared to vacuum fluctuations-based QRNGs while offering comparable generation rates in the order of $Gbps$. We first discuss the complete characterization of the subsystems in Sec.~\ref{sec:qrng_sub_systems}, followed by the development of the phase-noise-based QRNG system in Sec.~\ref{sec:PNB_QRNG}, and finally provide our conclusions in Sec.~\ref{sec:conclusion}. 

\section{QRNG SUBSYSTEMS CHARACTERIZATION}
\label{sec:qrng_sub_systems}

    The hardware components used to develop the quantum entropy source are characterized and tested to ensure accurate measurements and a comprehensive understanding of the system. The key subsystems include the laser source [Sec.~\ref{sub-sec:laser}], a fiber-based unbalanced Mach-Zehnder interferometer (uMZI) [Sec.~\ref{sub-sec:MZI}], a high-speed photodiode (PD) [Sec.~\ref{sub-sec:PD}], and the oscilloscope [Sec.~\ref{sub-sec:Oscilloscope}] used for measurements. 

    \subsection{Laser System}
    \label{sub-sec:laser}

    First, the average laser output power at various input voltages of the laser diode is measured to get the P-V characteristics of the laser system (WSLS-1550-005m-PM, Wave Spectrum). The laser’s average power output ($\mu W$) versus the input voltage ($mV$) is plotted to find the knee-point of the laser. The knee-point of the laser system used in the present experiment is found at 12.65 $\mu W$ of laser output power or 3.67 $mV$ of voltage provided to the laser diode.
    
    Next, the spectral width of the emission and the central wavelength of emission are observed on the YOKOGAWA AQ6374 Optical Spectrum Analyzer (OSA). The spectral line-width for the laser system is observed to be $\Delta \lambda$ = 0.0421 $nm$ or $\Delta \nu$ = 5.23 $GHz$. Likewise, the central wavelength of the laser system is observed to be $\lambda$ = 1551.1970 $nm$. The coherence time ($T_c$) of the laser is calculated based on the line-width measured. $T_c$ of the laser system is 0.19 $ns$. 
    
    \subsection{Unbalanced Mach-Zehnder Interferometer Design}
    \label{sub-sec:MZI}

\begin{figure} [ht]
       \begin{center}
            \begin{tabular}{c} 
                \includegraphics[height=5.5cm]{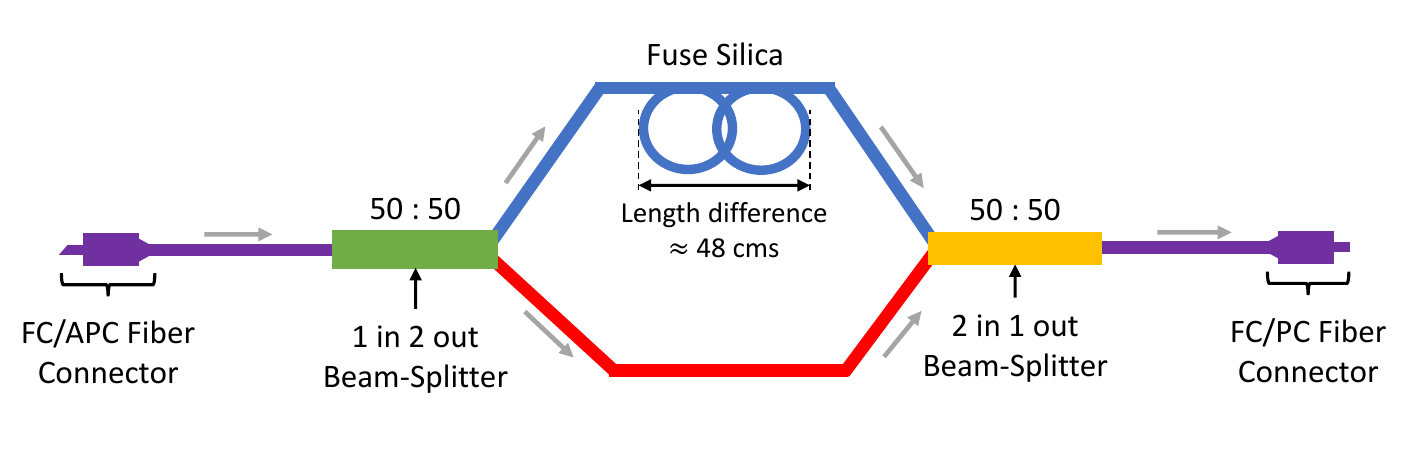}
            \end{tabular}
       \end{center}
       \caption[MZIDesign] 
       { \label{fig:MZIDesign} Unbalanced Mach-Zehnder Interferometer (uMZI) design.}
    \end{figure}

    A uMZI with an approximate delay length ($\Delta L$) of 48 $cm$ is used, resulting in a delay time ($T_d$) of 2.35 $ns$, which is much longer than the coherence time ($T_c$ = 0.19 $ns$) of the laser. The uMZI is configured as a single-input and single-output system, as depicted in Fig.~\ref{fig:MZIDesign} of the design schematic. The constructed fiber-based uMZI can be directly coupled with the laser's output via a fiber connector.

    \subsection{High-Speed Photodiode}
    \label{sub-sec:PD}

    An InGaAs 5.0 $GHz$ bandwidth photodiode (manufactured by Thorlabs, DET08CFC/M \cite{thorlabsinc_thorlabs}) is used to collect the photonic signal from the uMZI. The chosen photodiode has a wavelength response range between 800 $nm$ to 1700 $nm$, a maximum peak power of 100 $mW$, and a dark current of 10 $nA$.

    \subsection{Oscilloscope}
    \label{sub-sec:Oscilloscope}

    A 500 $MHz$ bandwidth (Techtronix \cite{Tektronix_MDO}) oscilloscope, with a max sampling rate of 2.5 $GS/sec$, is used to collect the electronic signal from the photodiode. The chosen oscilloscope has a maximum vertical resolution of 1 $mV/div$ with an in-built 8-bit Analog-to-Digital (ADC) converter. For QRNG hardware characterization and testing prior to packaging, the TekScope\cite{Tektronix_tekscope} software is used for control and continuous data acquisition from the oscilloscope.  

\section{DESIGN OF THE PHASE-NOISE-BASED QRNG}
\label{sec:PNB_QRNG}

    The inherent source of randomness for the phase-noise-based QRNG originates from the spontaneous emission of the laser. It is inherently random in nature, guaranteed by the origins of such randomness associated with vacuum fluctuations. The physical system of this QRNG is implemented by using a uMZI to perform a self-heterodyne measurement of the laser’s emission. The laser is operated close to its threshold in order to achieve the maximum possible quantum signal-to-classical noise ratio (QSCNR).

    When operated close to the threshold, the voltage output of the measurement system obtained on the oscilloscope from the photodiode\cite{qi_2010_highspeed} is given by:

    \begin{equation}
        V(t) \propto 2 E(t) E(t+\tau) Sin(\Delta\theta(t)) \propto P\Delta\theta(t)
        \label{eq:3}
    \end{equation}

    Where $\Delta\theta(t)$ is small such that $Sin(\Delta\theta(t)) \approx \Delta\theta(t)$, which is the relative phase between both the arms of the unbalanced interferometer given by $\Delta\theta(t) = \theta(t) - \theta(t+\tau)$. Here $\tau$ = 2.35 $ns$ as reported in Sec.~\ref{sub-sec:MZI}.

    To maximize the phase noise obtained and avoid any correlations, sampling rate matching is performed, detailed in Sec.~\ref{sub-sec:SamplingRateMatching}. Followed by QSCNR analysis in Sec.~\ref{sub-sec:QSCNR}. The same is verified by checking the RF Spectrum of the laser noise output to the phase noise output in Sec.~\ref{sub-sec:RFSpectrum}. Section~\ref{sub-sec: RawDataAcqNorm} deals with the process of data acquisition and normalization. Finally, Sec.~\ref{sub-sec:post_processing} details the post-processing, randomness testing, and system packaging of the QRNG. 

    \subsection{Sampling Rate Matching}
    \label{sub-sec:SamplingRateMatching}

        The spontaneous emission of a laser source cannot be associated with a specific rate of emission or generation, as it is completely random and governed by vacuum fluctuations. Therefore, for the phase-noise-based QRNG, we use the following experimental timing parameters to determine the effective generation rate of our system, as shown in Fig.~\ref{fig:HardwareTiming}. The parameters in consideration include:

        \begin{figure} [ht]
           \begin{center}
                \begin{tabular}{c} 
                    \includegraphics[height=5.5cm]{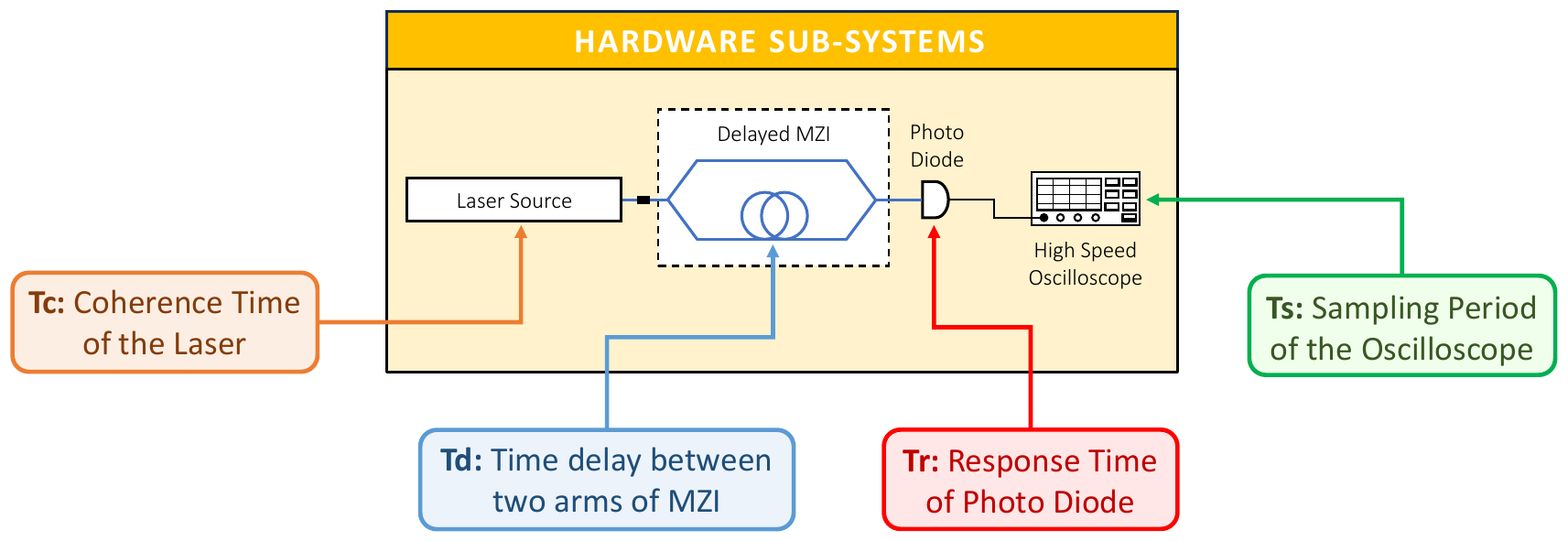}
                \end{tabular}
           \end{center}
           \caption[HardwareTiming] 
           { \label{fig:HardwareTiming} Hardware subsystem timing statistics. $T_c$ of the laser source, $T_d$ of the delayed MZI, $T_r$ of the photodiode, and $T_s$ of the oscilloscope.}
        \end{figure}

        \begin{enumerate}
        \item [$T_c$]: The coherence time of the laser, which is inversely proportional to the linewidth of the laser source, calculated to be 0.19~$ns$ as mentioned in Sec.~\ref{sub-sec:laser}.
            \item [$T_d$]: The time delay between both the arms of the uMZI, calculated to be 2.35 $ns$ as mentioned in Sec.~\ref{sub-sec:MZI}.
            \item [$T_r$]: The response time of the photodiode, which is inversely proportional to the small signal bandwidth $f_{-3dB}$\cite{thorlabs_2019_rise} of the photodiode.
                \begin{equation}
                    T_{r} = \frac{0.35}{f_{-3dB}} = \frac{0.35}{5} \; ns = 0.07 \; ns
                \end{equation}
            \item [$T_s$]: The sampling period of the oscilloscope, which is the reciprocal of the sampling rate of the oscilloscope. 
                    \begin{equation}
                        T_{s} = \frac{1}{f_{s}} = \frac{1}{0.25} \; ns = 4.00 \; ns
                    \end{equation}
        \end{enumerate}

        Based on the above-mentioned experimental parameters of the subsystems of the QRNG, we adhere to the following conditions to minimize correlations in the output obtained and decide the sampling rate of the oscilloscope for raw data acquisition\cite{qi_2010_highspeed}. 

         \begin{equation}
             T_d \gg T_c
             \label{eq:6}
         \end{equation}
            
         \begin{equation}
             T_s - T_d > T_r
             \label{eq:7}
         \end{equation}
    
        Equation~(\ref{eq:6}) is used to avoid any phase correlation in the system. Whereas, Eq.~(\ref{eq:7}) is used to reduce the correlation between adjacent samples due to uMZI length and the photodiode response time. The experimental timing parameters and the conditions for the current experimental system can be found in Table~\ref{tab:sampling_rate_conditions}.
    
        \begin{table}[ht]
            \caption{Sampling rate matching conditions for the phase-noise-based QRNG experimental system.}
            \label{tab:sampling_rate_conditions}
            \begin{center}
                \begin{tabular}{|l|l|l|l|l|l|}
                    \hline
                    \rule[-1ex]{0pt}{3.5ex} $T_c$ ($ns$) & $T_d$ ($ns$) & $T_r$ ($ns$) & $T_s$ ($ns$) & $T_d \gg T_c$ & $T_s - T_d > T_r$ \\
                    \hline
                    \rule[-1ex]{0pt}{3.5ex} 0.19 & 2.35 & 0.07 & 4.00 & True & True \\
                    \hline
                \end{tabular}
            \end{center}
        \end{table}

    \subsection{Phase Noise Measurement and Optimization of Quantum Noise\texorpdfstring{\cite{xu_2012_ultrafast}}{}}
    \label{sub-sec:QSCNR}

        The total fluctuation output from the system can be expressed as the sum of two components: (a) quantum phase fluctuations, which are inversely proportional to the laser output power and can be modeled as Gaussian white noise \cite{petermann_1988_laser}, and (b) classical phase noise, which is independent of the laser power \cite{vahala_1983_occupation}, as shown in Eq.~(\ref{eq:8}).
    
        \begin{equation}
            \langle \theta(t)^{2} \rangle = \frac{Q}{P} + C
            \label{eq:8}
        \end{equation}
    
        It is to be noted that the quantities $Q$, $P$, and $C$ are independent of time within the timeframe of measurement. Using Eq.~(\ref{eq:3}) and Eq.~(\ref{eq:8}) paired with the background of the photodiode $(F)$, the average variance of the output voltage $ \langle V(t)^{2} \rangle$ obtained from the photodiode on the oscilloscope can be described as:
    
        \begin{equation}
            \langle V(t)^{2} \rangle = ACP^{2} + AQP + F
            \label{eq:9}
        \end{equation}
    
        Where $\langle V(t)^{2} \rangle $ is the average voltage variance output from the photodiode, $ACP^{2}$ is the classical phase noise, $AQP$ is the quantum phase fluctuations, and $F$ is the background noise of the photodiode.
    
\begin{figure} [ht]
           \begin{center}
                \begin{tabular}{c} 
                    \includegraphics[height=16.5cm]{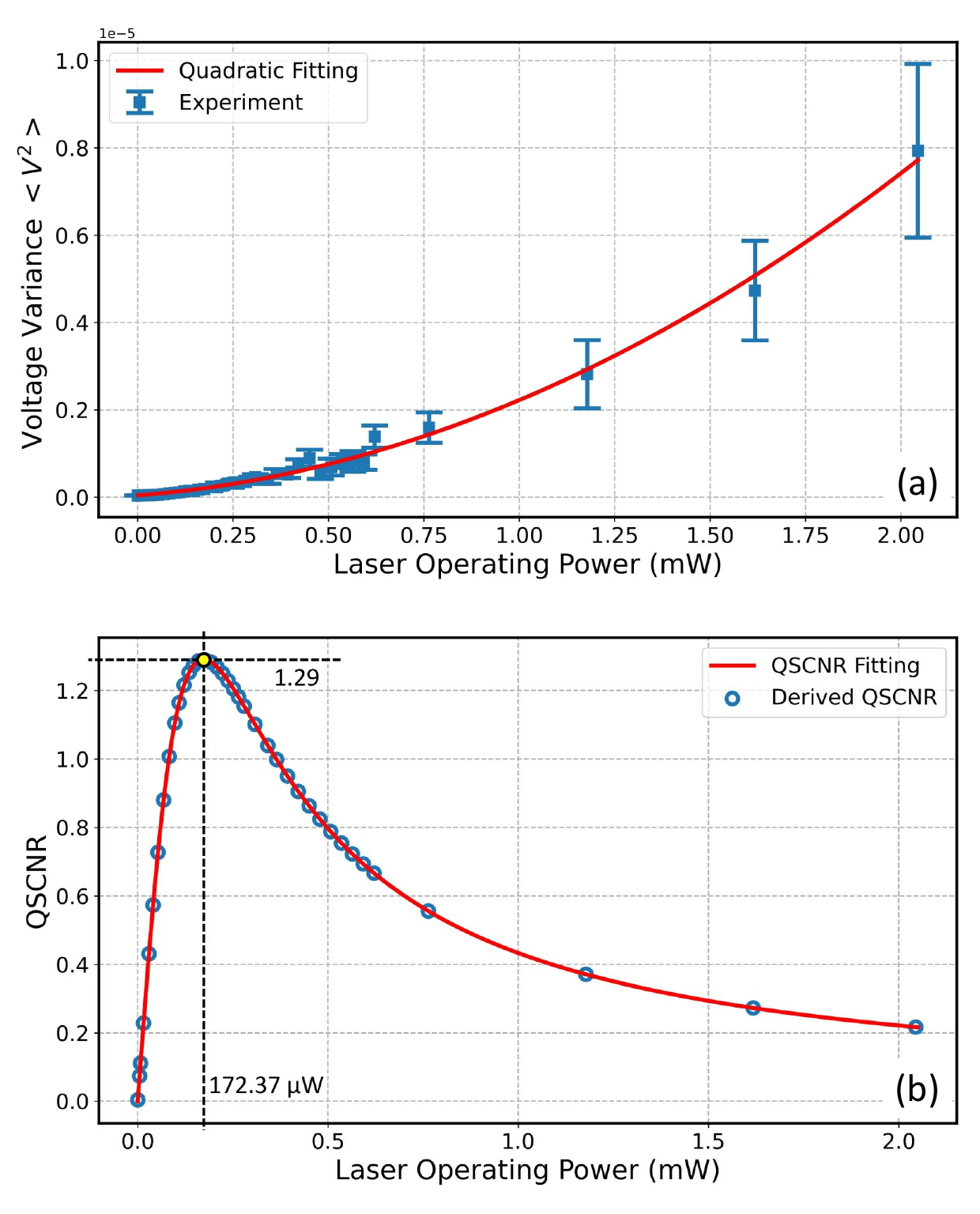}
                \end{tabular}
           \end{center}
           \caption[VoltageVarianceQSCNR] 
           { \label{fig:VoltageVarianceQSCNR} Voltage variance and QSCNR analysis: (a) The laser operating power $(mW)$ on x-axis and the voltage variance $ \langle V^{2} \rangle$ of laser output on y-axis, and (b) The laser operating power in $mW$ on the x-axis, and the quantum signal to classical noise ratio ($QSCNR$) on the y-axis.}
        \end{figure}

        We experimentally measure the average voltage variance output by collecting 10 consecutive datasets ($\approx$ 3 $Gb$ raw oscilloscope data) of the voltage output from the photodiode for 39 different laser power outputs, approximately ranging from 0 $mW$ to 2 $mW$ ($\approx$ 117 $Gb$ raw oscilloscope data) as shown in Fig.~\ref{fig:VoltageVarianceQSCNR}(a).  
    
        We fit a parabola or quadratic polynomial function and estimate the parameters $AQ$, $AC$, and $F$ from Eq.~(\ref{eq:9}), the values of which are shown in Table \ref{tab:parameter_values_eq9}. 
    
        \begin{table}[ht]
        \caption{Parameter values of Eq.~(\ref{eq:9}).}
        \label{tab:parameter_values_eq9}
        \begin{center}
        \begin{tabular}{|l|l|}
        \hline
        \rule[-1ex]{0pt}{3.5ex} Parameter & Value \\
        \hline
        \rule[-1ex]{0pt}{3.5ex} $AC$ ($mV^{2}/mW^{2}$) & $1.51 \times 10^{-6}$ \\
        \hline
        \rule[-1ex]{0pt}{3.5ex} $AQ$ ($mV^{2}/mW$) & $6.72 \times 10^{-7}$ \\
        \hline
        \rule[-1ex]{0pt}{3.5ex} $F$ ($mV^{2}$) & $4.50 \times 10^{-8}$ \\
        \hline
        \end{tabular}
        \end{center}
        \end{table}

        To obtain maximum quantum phase fluctuations, we devise a QSCNR from Eq.~(\ref{eq:9}) as follows:
    
        \begin{equation}
            QSCNR = \frac{AQP}{ACP^{2}+F}
        \end{equation}
    
        Given the experimental parameters $AQ$, $AC$, and $F$, a suitable laser output power is found such that we maximize the value of QSCNR to harvest maximum quantum phase fluctuations from the hardware system, as shown in Fig.~\ref{fig:VoltageVarianceQSCNR}(b).
    
        The highest measured QSCNR is 1.29 at 172.37 $\mu W$ of laser power output. Therefore, 172.37 $\mu W$ output power or the corresponding 4.9 $mV$ of input voltage given to the laser is the optimal point of operation of the laser to harvest maximum quantum phase fluctuations.

    \subsection{RF Spectrum Analysis \texorpdfstring{\cite{xu_2012_ultrafast}}{}}
    \label{sub-sec:RFSpectrum}

    To check for the effects of the phase noise as compared to the intensity noise of the laser, we check the RF Spectrum with (a) The background of the RF Spectrum Analyzer (b) The photodiode's background by connecting the laser (turned off) with the PD and analyzing the RF spectrum (c) The laser intensity noise at 172.37 $\mu W$ power output (d) The phase noise at 172.37 $\mu W$ power output of the laser analyzed by checking the RF Spectrum connecting the laser with the uMZI followed by the PD.
    
\begin{figure} [ht]
           \begin{center}
                \begin{tabular}{c} 
                    \includegraphics[height=8cm]{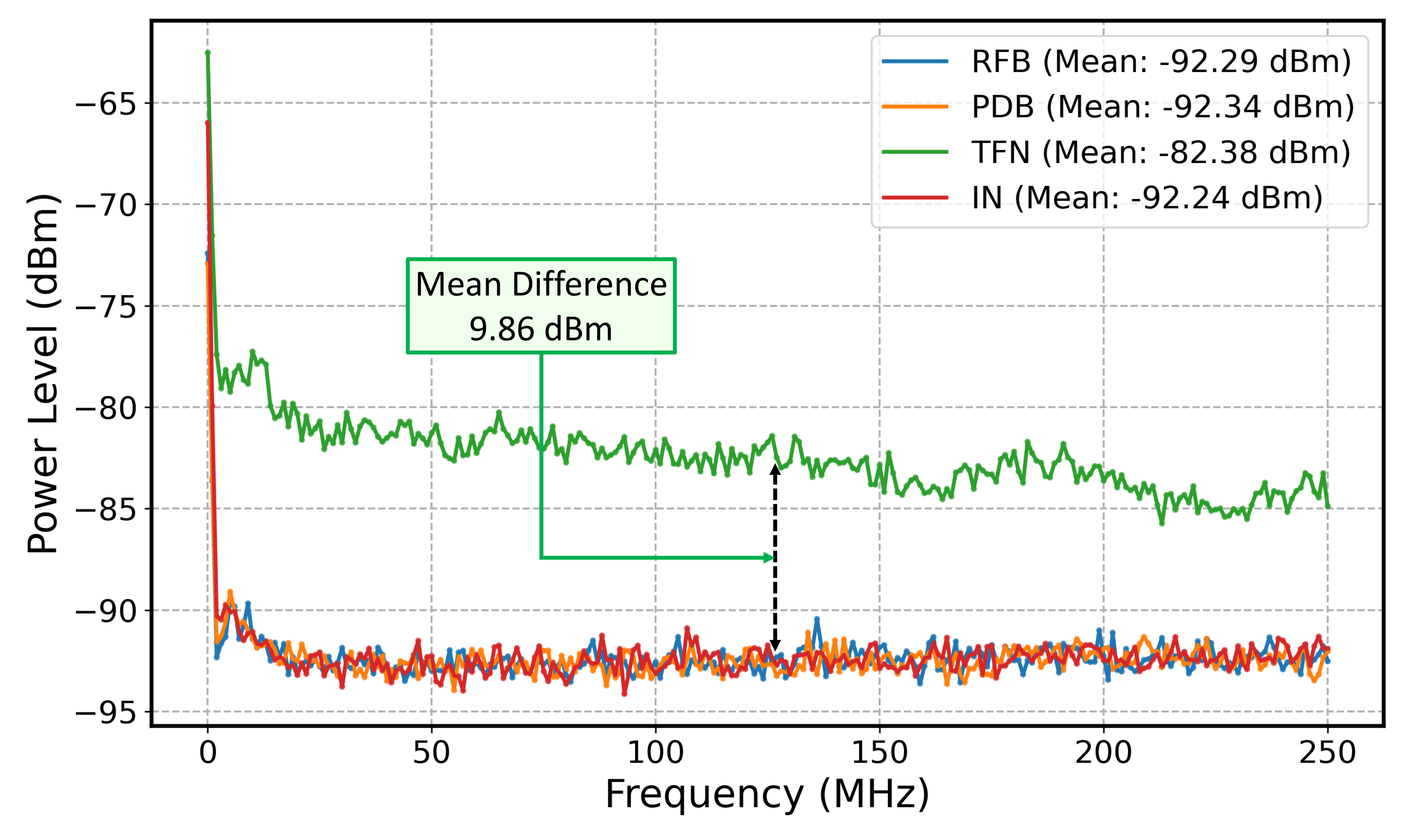}
                \end{tabular}
           \end{center}
           \caption[RFSpectrum] 
           { \label{fig:RFSpectrum} RF spectrum analysis of the system where $RFB$ is the RF spectrum analyzer background, $PDB$ is the photodiode's background, $TFN$ is the total fluctuation noise of the system at 172.37 $\mu W$ power output of the laser, and $IN$ is the intensity noise of the laser operated at 172.37 $\mu W$ of power output.}
        \end{figure}

    We observe a difference in the average phase noise ($TFN$) of 9.86 $dBm$ compared to the intensity noise of the laser ($IN$) and the standard backgrounds ($RFB$, $PDB$) as shown in Fig.~\ref{fig:RFSpectrum}, verifying our claim of working with high $QSCNR$. 

    \subsection{ Raw Data Acquisition and Normalization}
    \label{sub-sec: RawDataAcqNorm}

        Based on the sampling rate matching conditions specified in Sec.~\ref{sub-sec:SamplingRateMatching}, the sampling rate of our oscilloscope $(f_{s})$ is set at 250 $MS/sec$. Given the oscilloscope's ADC resolution $(\alpha_{res})$ at 8-bits, the generation rate for the QRNG is defined as:

        \begin{equation}
            \xi_{raw} = f_{s} \times \alpha_{res}
            \label{eq:11}
        \end{equation}
        
        Where $\xi_{raw}$ is the QRNG generation rate in bits/second ($bps$), $f_{s}$ is the sampling rate of the oscilloscope in samples/second ($Sps$), and $\alpha_{res}$ is the resolution of the ADC in bits ($b$). Therefore, based on Eq.~(\ref{eq:11}), the raw data generation rate $(\xi_{raw})$ of the phase-noise-based QRNG is 2.0 $Gbps$. 
        
        The voltage output obtained from the oscilloscope consists of floating-point signed decimal data, which is normalized to unsigned 8-bit binary data by using:

        \begin{equation}
            \beta_{i} = \frac{\delta_i - \delta_{min}}{\delta_{max} - \delta_{min}} \times (2^{\alpha_{res}} - 1), \,\,\,i \in [0,255]
            \label{eq:12}
        \end{equation}

        Where, $\beta_{i}$ is the $i^{th}$ unsigned binary bit, $\delta_{i}$ is the $i^{th}$ signed floating point decimal, $\delta_{min}$ is the minimum signed floating point decimal of the given data, $\delta_{max}$ is the maximum signed floating point decimal of the given data, and $\alpha_{res}$ is the resolution of the ADC ($b$).

\begin{figure} [ht]
           \begin{center}
                \begin{tabular}{c} 
                    \includegraphics[height=4cm]{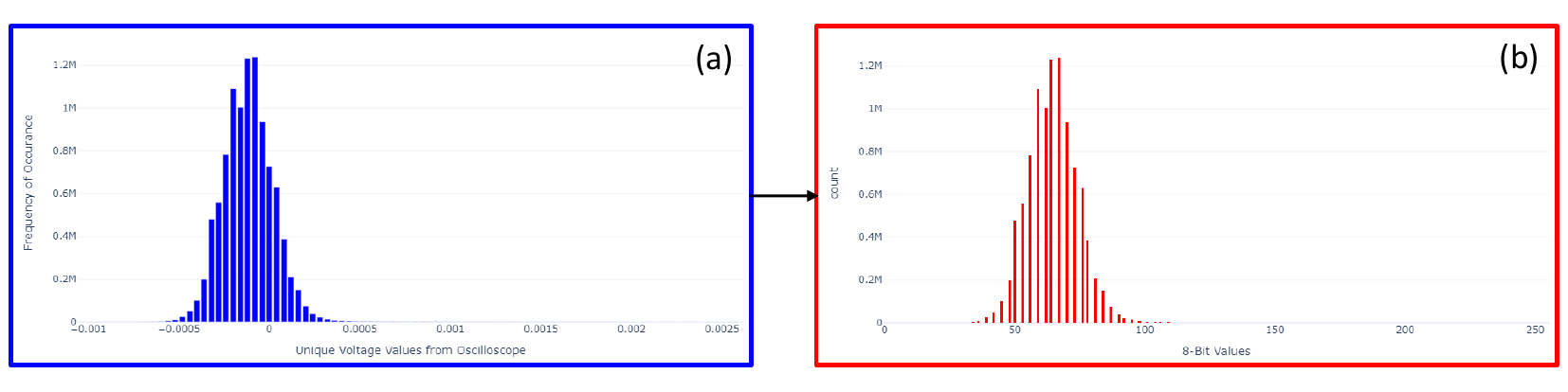}
                \end{tabular}
           \end{center}
           \caption[ADCNorm] 
           { \label{fig:ADCNorm} Frequency distributions of: (a) Raw data obtained from oscilloscope (blue) and (b) Normalized 8-bit data (red).}
        \end{figure}

        Initially, 30 sets of raw data (2.4 $Gb$ of binary bits) are acquired from the hardware system. The frequency distribution plot of the output, and the 8-bit normalized data plot of the output, as shown in Fig.~\ref{fig:ADCNorm}. A lack of complete binning of the discrete $2^{8}$ value possible with an 8-bit ADC is observed due to the small signal strength from the interferometer and the lack of better vertical resolution of the oscilloscope over the current maximum at 1 $mV/div$. 

    \subsection{Post-Processing, Validation and System Packaging}
    \label{sub-sec:post_processing}
        After acquiring the 8-bit normalized raw data, a TSE-based post-processing framework~\cite{aksv_2024_tsefpga} is employed to perform randomness extraction using FPGA systems. The min-entropy of the raw data is 4.15/8~$bits$ (i.e., 51.875\%)~\cite{ksv_2024_tsecpugpu}. Based on this, the extraction ratio $(\eta_{er})$ is set to 0.5 for FPGA-based post-processing~\cite{aksv_2024_tsefpga}. Given $\eta_{er}$ and the raw generation rate $\xi_{raw}$, the post-processed generation rate of the QRNG $(\xi_{pp})$ can be expressed as:
        \begin{equation} 
            \xi_{pp} = \eta_{er} \times \xi_{raw}
        \label{eq:QRNG_Post-Processed_Speed}
        \end{equation}
        Therefore, the effective $\xi_{pp}$ of the phase-noise-based QRNG is 1~$Gbps$ based on Eq.~(\ref{eq:QRNG_Post-Processed_Speed}). By carefully engineering the hardware architecture, the QRNG system can achieve near–raw generation rates after post-processing. The post-processed data are subsequently validated using standard statistical randomness test suites such as NIST and Diehard. The raw data exhibit weaknesses and fail the majority of these tests, whereas the post-processed data successfully pass all of them with standard statistical confidence levels~\cite{ksv_2024_tsecpugpu}.
                
        Following successful validation, the complete QRNG system is packaged for real-time operation. In this implementation, the oscilloscope used during testing is replaced with an ADC interfaced to an FPGA for on-board post-processing~\cite{aksv_2024_tsefpga}, as described in Fig.~\ref{fig:PackagedPhaseNoiseQRNG}. The resulting packaged QRNG device consists of a laser integrated with a uMZI and PD as the entropy source, followed by an ADC-FPGA chain that provides real-time post-processed random number output via USB for demonstration purposes.
                
        It is to be noted that the achievable generation rate of the QRNG is constrained by the FPGA's output interface. Among the commonly available interfaces, PCIe and SFP offer the highest data throughput suitable for sustained multi-gigabit streaming, while Ethernet provides moderate speeds, and USB represents the slowest link. The developed QRNG device has been evaluated under controlled laboratory and live demonstration conditions, exhibiting consistent results that confirm its robustness and operational stability. These findings indicate near-deployable maturity corresponding to TRL 7 and approaching TRL 8.

\section{CONCLUSION AND FUTURE SCOPE}
\label{sec:conclusion}

        \begin{figure} [ht]
           \begin{center}
                \begin{tabular}{c} 
                    \includegraphics[height=7.5cm]{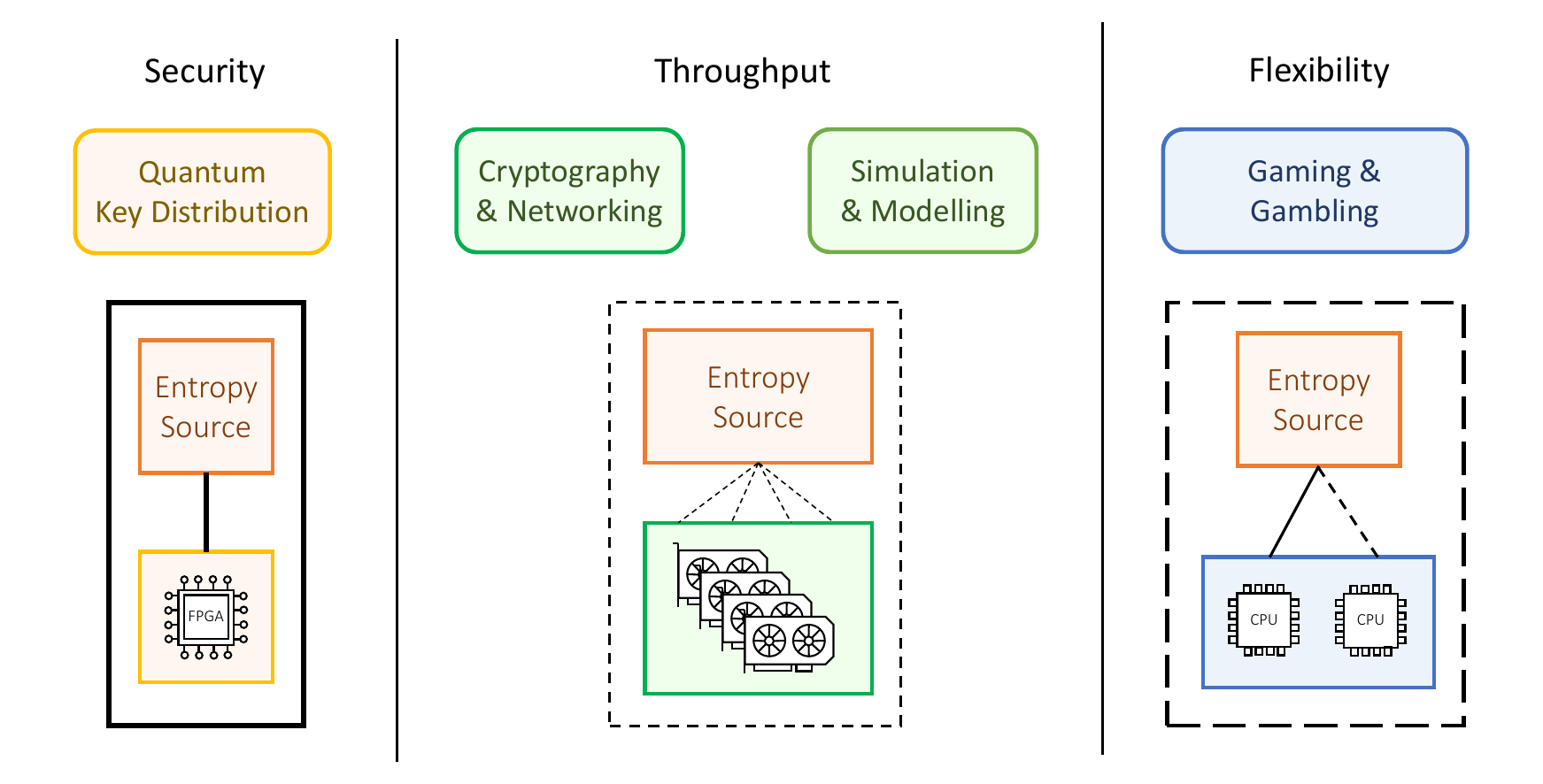}
                \end{tabular}
           \end{center}
           \caption[applications] 
           { \label{fig:applications} Application specific deployment of QRNG systems.}
        \end{figure}
        
     This work presents the end-to-end hardware development of a phase-noise-based QRNG system, spanning subsystem characterization to final system packaging, resulting in a TRL 7 system demonstrated in an operational environment. Future work focuses on qualification, interface standardization, and deployment-scale validation towards achieving TRL 8 product status. For scalability in terms of generation rates, the subsystem sampling rate matching metrics can be tuned\cite{anuragksv_2024_qrng}. The robustness and quality of the phase-noise-based QRNG can be further improved by building an ultra-stable Planar Lightwave Circuit (PLC) based delayed uMZI with precise temperature control \cite{xu_2012_ultrafast}. Similarly, on-chip realization of the phase-noise-based QRNG can also be made via leveraging various hybrid packaging techniques \cite{Huang_2025_onchipphasenoiseqrng}.

     The current implementation of the phase-noise-based QRNG with FPGA for post-processing is well-suited for absolute security, sealed-box applications such as QKD systems, as well as for use in the defence industry. In contrast, applications including networking, modeling, and simulation emphasize higher random number generation throughput. For these applications, we can make use of QRNGs connected to a distributed GPU network for multifold improvement in extraction speeds as compared to shown in this work. Finally, a more flexible approach may be employed for standalone, application-specific devices by pairing the QRNG’s entropy source with single or multiple CPUs, depending on efficiency, demand, and resource utilization. Such configurations are beneficial for gaming, gambling, and general-purpose applications on mobile devices, arcade systems, and similar platforms. Therefore, for the broader industry-adoption of QRNG technology, we envision application-specific QRNG deployments featuring different form-factors of entropy sources integrated with different computational hardware to perform post-processing\cite{qiao_2025_realtimeonchiphomodyneqrng, Huang_2025_onchipphasenoiseqrng, Crampton_2025_lowswapphqrng}, as illustrated in Fig.~\ref{fig:applications}.

\acknowledgments

    The authors acknowledge the support of DIAT (DU) under the Grant-in-Aid program. They express their sincere gratitude to Prof. C. S. Unnikrishnan and the late Dr. Sankaranarayanan Selvarajan of the School of Quantum Technology, DIAT, for their assistance with laboratory work, valuable discussions, and insightful feedback.

\bibliography{report}
\bibliographystyle{spiebib}

\end{document}